\documentclass[preprint]{elsarticle}
\usepackage[x11names]{xcolor}
\usepackage{color, colortbl}
\usepackage[utf8]{inputenc}
\usepackage{float} 
\usepackage{nicematrix}
\usepackage{adjustbox}
\usepackage{multirow}
\usepackage{graphicx, subcaption}
\usepackage{hyperref}
\DeclareUnicodeCharacter{0301}{\'{e}}
\usepackage{multirow}
\usepackage{textgreek}
\usepackage{cleveref}
\usepackage{pdfpages}
\usepackage{hyperref}

\usepackage[section]{placeins}
\usepackage{changepage}
\usepackage{svg}

\journal{Applied Acoustics}

\begin{document}

\begin{frontmatter}

% \title{Foundation modelling for Automatic Ship-Radiated Noise Monitoring using Contrastive Learning.}
\title{The Computation of Generalized Embeddings for Underwater Acoustic Target Recognition using Contrastive Learning}

\author[inst1]{Hilde I. Hummel*}
\ead{h.i.hummel@cwi.nl}
\affiliation[inst1]{organization={Centre of Mathematics and Computer Science, Department of Stochastics},%Department and Organization
            addressline={Science Park 123}, 
            city={Amsterdam},
            postcode={1098 XG},
            country={Netherlands}}

\author[inst1]{Arwin Gansekoele}
\author[inst2]{Sandjai Bhulai}
\author[inst1,inst2]{Rob van der Mei}

\affiliation[inst2]{organization={Vrije Universiteit, Department Mathematics},%Department and Organization
            addressline={De Boelelaan 1111}, 
            city={Amsterdam},
            postcode={1081 HV},
            country={Netherlands}}

\cortext[inst1]{Corresponding author}

\begin{abstract}
%% Text of abstract
The increasing level of sound pollution in marine environments poses an increased threat to ocean health, making it crucial to monitor underwater noise. By monitoring this noise, the sources responsible for this pollution can be mapped. Monitoring is performed by passively listening to these sounds. This generates a large amount of data records, capturing a mix of sound sources such as ship activities and marine mammal vocalizations. Although machine learning offers a promising solution for automatic sound classification, current state-of-the-art methods implement supervised learning. This requires a large amount of high-quality labeled data that is not publicly available. In contrast, a massive amount of lower-quality unlabeled data is publicly available, offering the opportunity to explore unsupervised learning techniques. This research explores this possibility by implementing an unsupervised Contrastive Learning approach. Here, a Conformer-based encoder is optimized by the so-called \textit{Variance-Invariance-Covariance Regularization} loss function on these lower-quality unlabeled data and the translation to the labeled data is made. Through classification tasks involving recognizing ship types and marine mammal vocalizations, our method demonstrates to produce robust and generalized embeddings. \textcolor{black}{This shows to potential of unsupervised methods for various automatic underwater acoustic analysis tasks.}
 
\end{abstract}

%%Graphical abstract
\begin{graphicalabstract}
\includegraphics{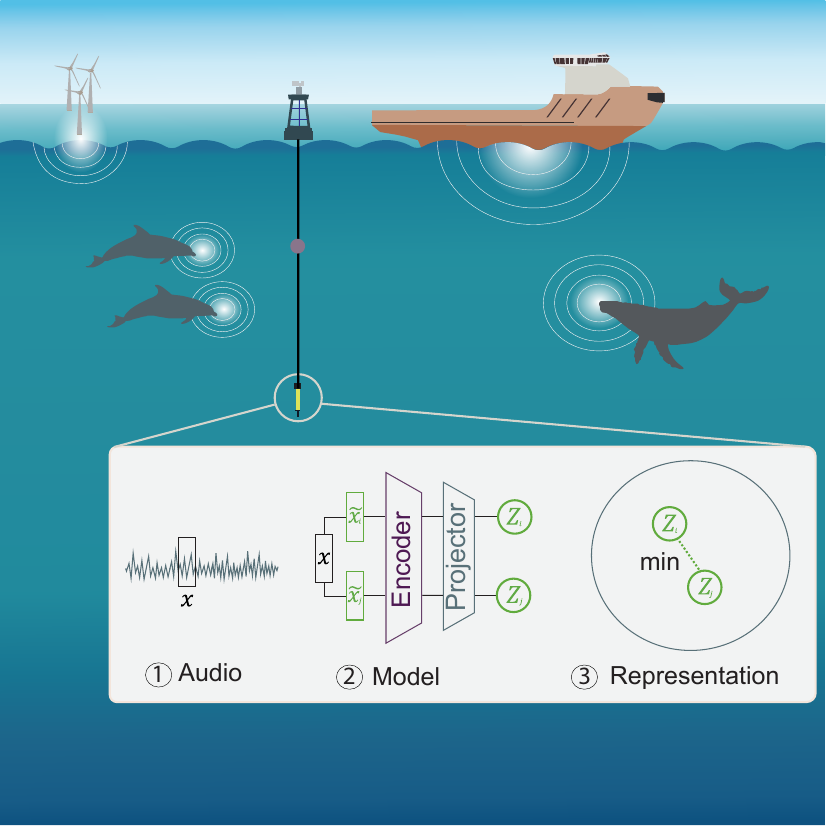}
\end{graphicalabstract}

%%Research highlights
\begin{highlights}
\item \textcolor{black}{An application of unsupervised contrastive learning (CL) on unlabeled underwater acoustic datasets for Underwater Acoustic Target Recognition. It demonstrates a novel implementation on abundantly available unlabeled data collected from a single hydrophone for the first time.}
\item \textcolor{black}{This novel framework effectively translates unstructured acoustic data into generalized embeddings which can be used for various downstream classification tasks.}
\item \textcolor{black}{The proposed unsupervised CL framework produces robust and generalized embeddings. This creates the potential for this framework to translate to other underwater acoustic analysis tasks, such as climate change monitoring and nuclear bomb test detection.}
\end{highlights}

\begin{keyword}
%% keywords here, in the form: keyword \sep keyword
Ship Radiated Noise \sep Unsupervised Learning \sep Contrastive Learning \sep Underwater Sound \sep Marine Mammals \sep UATR
%% PACS codes here, in the form: \PACS code \sep code
\PACS 0000 \sep 1111
%% MSC codes here, in the form: \MSC code \sep code
%% or \MSC[2008] code \sep code (2000 is the default)
\MSC 0000 \sep 1111
\end{keyword}

\end{frontmatter}

%% main text
\section{Introduction}

The ocean environment is polluted by the increased level of artificially induced sounds. \textcolor{black}{This negatively impacts the ocean health directly.} For this reason, monitoring underwater acoustics is crucial to protect our ocean environments. \textcolor{black}{By monitoring the underwater sound, the sources of noise responsible for pollution can be mapped. }One such system to monitor ocean sound is the implementation of a passive acoustic monitoring system. Here, hydrophones are placed to passively listen to the underwater acoustics. These acoustics originate from multiple sources, such as ships, marine mammals, surface waves, and rain. Due to the mix of acoustic sources and complexity of the acoustic ocean environment, the analysis of underwater recordings becomes challenging. In addition, these systems generate a large amount of data, creating the need for automatic classification of underwater acoustic sounds, called \textit{underwater acoustic target recognition} (UATR). To automate UATR, machine learning has shown its potential \citep{thomas2020marine, hamard2024deep, hummel2024survey, licciardi2024whalenet}. 

Until now, the development of UATR algorithms has focused on supervised learning, which requires a vast amount of high-quality labeled acoustic data for training. Unfortunately, the amount of publicly available labeled acoustic data is limited. In contrast, a large amount of unlabeled underwater acoustic data is available to the public. To the best of our knowledge, no research has been conducted on the application of these lower-quality data utilizing  unsupervised learning models for UATR. One promising unsupervised approach to exploit these unlabeled acoustic data is Contrastive Learning (CL). This method tries to maximize the similarity between positive data samples, while minimizing the similarity for unrelated data samples. Within the computer vision domain, several CL methods are proposed to achieve remarkable performance using unlabeled datasets, such as SimCLR \citep{uid} and Momentum Contrast (MoCo) \citep{he2020momentum}.

In addition to computer vision tasks, this specific method has excellent performance in general audio tasks, as seen in \citep{saeed2021} and \citep{fonseca2021unsupervised, niizumi2022byol}. Furthermore, multiple studies have already implemented a form of CL to recognize the unique acoustic signatures of ships. For example, \citep{nie2023contrastive} proposed a ResNet-based encoder and optimized this in a supervised setting. In this study, they forced the recordings of the same ship to be close together in the embedding space while pushing the recordings of other vessels further away. They built a distance-based classifier on top of their CL framework to avoid overfitting under limited data. Similarly, \citep{sun2023underwater} proposed a three-layer \textit{Multi-Layer Perceptron} (MLP) optimized using the supervised contrastive loss function \citep{khosla2020supervised}. They did not include any augmentations and only used the label information, achieving over 98\% accuracy on ship type classification. Another supervised CL framework for ship type classification is proposed in \citep{xie2023guiding}. In this study, they proposed two separate ResNet encoders, where both encoders process two different transformations of the same audio. They implemented a combined loss of cosine similarity and the cross-entropy loss, achieving 80\% accuracy. In addition to these supervised methods, \citep{tian2023few} proposed a few-shot learning method to cope with limited labeled data. In this study, they proposed a four-stage training scheme inspired by SimCLR \citep{uid} and BYOL \citep{grill2020bootstrap}. They treated a dataset of labeled ship recordings as partially unlabeled and explored the impact of reducing the number of labeled data samples on their proposed method. In this study, they showed that they only needed 10\% of the labels to achieve the desired performance. A similar few-shot learning method is proposed in \citep{lin2024underwater} using a MoCo-based framework to achieve over 96\% accuracy. 

Although prior UATR studies report high recognition accuracy scores, their ability to generalize to similar tasks remains unexplored \citep{muller2024navigating}. \textcolor{black}{Generalization of models is needed to transfer the knowledge captured within the model to related tasks, making the generalized model applicable to various UATR tasks. It is an important property to enable more broadly applicable UATR.} This study aims to address this limitation by generating generalized embeddings from publicly available unlabeled data. These embeddings are optimized by the implementation of CL. The ability of the model to generalize is tested by implementing a simple classifier on Deepship \citep{irfan2021deepship}, ShipsEar \citep{santos2016shipsear} for ship type classification, and The Best of Watkin's \citep{sayigh2016watkins} for marine mammal sound classification. 
\textcolor{black}{The contribution of this paper is three-fold:}

\begin{itemize}
    \item The first implementation of unsupervised CL on a separate unlabeled underwater acoustic dataset is presented.
    \item A translation from this raw unlabeled data from a single hydrophone to a labeled classification task is made.
    \item The unsupervised CL approach is shown to generate robust and generalized embeddings.
\end{itemize}

\textcolor{black}{Altogether, this work shows the potential of an unsupervised CL framework for automatic UATR trained on highly available unlabeled data. This removes
the dependency on a vast amount of high-quality labeled data and can be applied
to various UATR tasks. Furthermore, this proof-of-concept can be extended by including more freely available unlabeled underwater acoustic data and by introducing more underwater acoustic analysis tasks.}

\section{Material and Methods}

\begin{figure}
    \centering
    \includegraphics[width=0.9\textwidth]{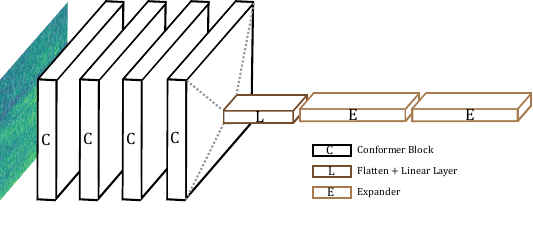}
    \caption{The proposed architecture for the unsupervised framework. The framework starts with a Mel spectrogram, followed by four Conformer Blocks (indicated in black), one flatten layer combined with a linear layer (indicated in brown), and finalized with an Expander (indicated in light brown).}
    \label{fig:Pipeline}
\end{figure}

A CL pipeline is proposed to generate generalized and informative embeddings, where the proposed architecture is visualized in Figure~\ref{fig:Pipeline}. In this architecture, an encoder is defined to compute the embeddings, and an expander is added to project these embeddings to a higher-dimensional space for optimization. After training this pipeline, the generated embeddings are used to fit simple task-specific classification models.

\subsection{Data}
In this work, both labeled and unlabeled data are utilized to train and evaluate the model. 

\subsubsection{Unlabeled datasets}
For the model's training, a single hydrophone from the Ocean Network Canada (ONC) dataset \citep{ONCurlPortal} was selected. The hydrophone \footnote{\url{https://data.oceannetworks.ca/DataSearch?locationCode=SCVIP&deviceCategoryCode=HYDROPHONE}} is located in the bay near Vancouver and is characterized by abundant shipping activity (longitude -123.43 West and latitude 49.04 North at 298 m depth). This specific hydrophone was selected to align with the same site as the Deepship benchmark dataset. Therefore, the characteristics of the dataset are similar to those of the Deepship data. From the recordings of this selected hydrophone, multiple 5-minute recordings were selected within the time range of September 2019 to December 2020. To ensure comprehensive representation, the selection procedure ensured equal representation of every month and time of day (morning, afternoon, evening, and night), resulting in 30 GB of raw data. Together, no information on the content of these data is available, making the implementation of CL challenging (see section \ref{section:loss_function}).

\subsubsection{Labeled datasets}
For the evaluation of the proposed method, three labeled datasets were selected as benchmark. \textcolor{black}{These datasets are widely used in related research, making it easier to compare the results with previous studies \citep{hummel2024survey}.} Each benchmark offers a different geographical area and labels, summarized as follows:
\begin{enumerate}
    \item \textbf{Deepship} \citep{irfan2021deepship}: This benchmark dataset contains 47 hours and 3 minutes of single-ship recordings categorized into four classes. The recordings were made in the strait of Georgia near Vancouver.
    \item \textbf{ShipsEar} \citep{santos2016shipsear}:  A smaller dataset recorded near Port Ria de Vigo, totaling 3 hours and 8 minutes of single-ship recordings categorized into five classes.
    \item \textbf{The Best of Watkin's} \citep{sayigh2016watkins}: A marine mammal vocalization dataset categorized into 45 classes. All recordings were made over a span of seven decades in a wide range of oceanic areas.
\end{enumerate}

The Deepship dataset was split into a training and test set using a time-wise split, ensuring that the training set represents the past and the test set represents the future. The recordings until November 2017 were used as training data and the recordings from December 2017 onward as test set. A time-wise split ensures that the proposed framework is most representative of practical systems, \textcolor{black}{making a better evaluation for real-world deployment}. However, due to this split, almost 60\% of the data samples in the test set are newly seen ships. \textcolor{black}{This requires the model to learn generalizable embeddings to correctly classify these previously unseen ships. This property is crucial for generalized models, as they need to be applicable to various downstream tasks and therefore robust to distribution shifts \citep{van2024birds}. }The remaining labeled datasets had no available time information and therefore were divided into 80\% training data and 20\% test data. 

\subsubsection{Preprocessing}
After the datasets were split, the audio samples were downsampled to 16 kHz. Doing so reduces the dimensionality of the data while minimizing the loss of important information captured within the audio. From these downsampled audio, nonoverlapping two-second windows were created.

\subsection{Augmentations}
\begin{figure}
    \centering
    \begin{subfigure}[b]{0.9\textwidth}
    \centering
    \includegraphics[width=0.5\textwidth]{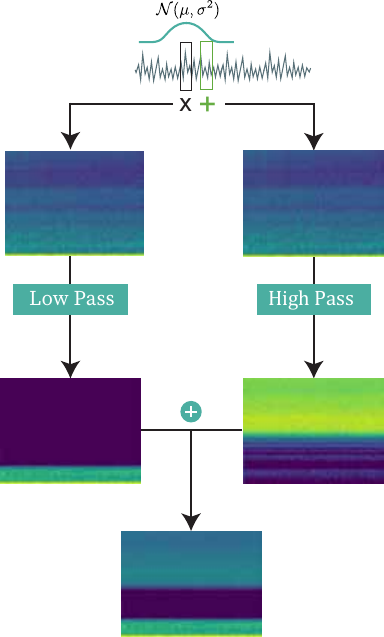}
    \caption{The MixUp augmentation pipeline.}
    \label{fig:MixUp}
    \end{subfigure}
    \begin{subfigure}[b]{0.9\textwidth}
        \includegraphics[width=1\linewidth]{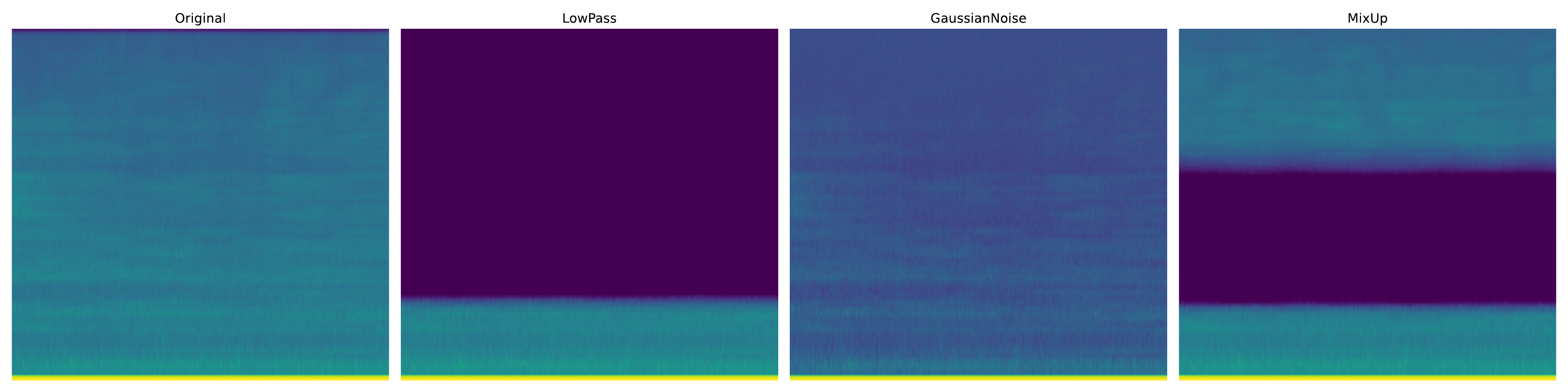}
    \caption{Visualization fo the proposed augmentation functions in the augmentation family.}
    \label{fig:AugmentedSpecs}
    \end{subfigure}
    \caption{Visualization of the augmentation family proposed for defining positive samples during training. The top image shows the pipeline for the MixUp strategy and the down image shows the resulting spectrograms for all augmentation functions.}
\end{figure}

In a CL framework, it is necessary to define positive samples for training. Since no information on the unlabeled training data is available, the audio was augmented to generate positive samples. Here, these positive samples are defined by applying two individual augmentation functions to a single audio fragment in the time domain. These two augmentation functions are randomly chosen from a family of augmentation functions. This family consists of four different augmentation functions, namely:

\begin{enumerate}
    \item \textbf{No augmentation:} Retains the original audio.
    \item \textbf{Gaussian noise:} Adds random Gaussian noise and constrains the output SNR between 0.3 dB and 0.5 dB. \textcolor{black}{The addition of Gaussian noise simulates more noisy different underwater acoustic environments.}
    \item \textbf{Low-pass filter:} Filters the input audio by a maximum of 1 kHz as the cutoff frequency. As stated in \citep{hummel2024survey}, the most discriminative information for the classification of the type of ships lies between 50 Hz and 1 kHz. \textcolor{black}{Therefore, a low-pass filter is a representative augmentation function.}
    \item \textbf{Mixup strategy:} Combines two filtered audio samples \textcolor{black}{and is inspired by the work of \citep{tonekaboni2021unsupervised} who presented \textit{Temporal Neighborhood Coding} for positive sample definition in time series data}. The original audio sample is filtered by the 1 kHz low-pass filter as previously described. These lower frequency bands capture the most unique acoustic ship characteristics \citep{hummel2024survey}. A second audio sample is selected by normal distribution over time, using the center time of the recording as $\mu$ and 50 seconds as $\sigma$. From this distribution, another two-second audio sample is drawn. This drawn sample is then filtered by a high-pass filter from 1 kHz-8 kHz, capturing related but different high-frequency band characteristics. Finally, these filtered samples are added, forming a mixed audio sample. \textcolor{black}{A MixUp strategy is commonly applied in audio pipelines \citep{kahl2021birdnet, xu2023self}, this adds more contrast to lower-contrast spectrograms.} The complete mixup pipeline is illustrated in Figure \ref{fig:MixUp}.
\end{enumerate}
The proposed augmentation functions are visualized in Figure \ref{fig:AugmentedSpecs}.

\subsection{Conversion to Mel spectrogram}
The augmented audio is converted to a Mel spectrogram. This is a time-frequency representation of the time-domain audio convolved by the Mel scale. This Mel scale comprises multiple half-overlapping triangular filters designed to mimic the human perception of sound; in this study the number of Mel filters was set to 128. Each triangular filter is centered at frequency $f_{mel}$ which is defined as:
\begin{equation}
    f_{mel} (Hz) = 2595 \times\log_{10}\Big(1+\frac{f (Hz)}{700}\Big).
\end{equation}

\subsection{Architecture}
\begin{figure}
    \centering
    \includegraphics[width=1\textwidth]{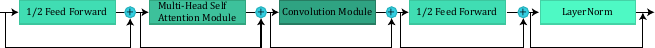}
    \caption{A single Conformer Block.}
    \label{fig:ConformerBlock}
\end{figure}

The architecture of the proposed method combines an encoder with an expander (see Figure \ref{fig:Pipeline}). In this research, a small architecture is chosen to create a proof of concept. In this regard, the encoder is defined by four Conformer blocks \citep{gulati2020conformer}, each including feedforward layers, four multi-head self-attention heads, and convolutional modules with a kernel size of 31 (see Figure \ref{fig:ConformerBlock}). \textcolor{black}{Previous research has shown the potential of the Conformer model in audio classification on AudioSet \citep{srivastava2022conformer}.} The resulting embeddings are flattened to a 2,048-size vector and passed through a linear layer. The architecture is finalized by the expander, composed of a two-layer MLP of dimension 8,196.

\subsection{Loss function}\label{section:loss_function}
The unlabeled training dataset can be viewed as a single long recording. Since there is no information on the content of this recording available, it is difficult to assign negative samples as proposed in the original SimCLR paper \citep{uid}. For this reason, the \textit{Variance-Invariance Covariance Regularization} (VICReg)\citep{bardes2021vicreg} loss function is implemented. Here, no prior definition of negative samples is needed. The loss function consists of three components: the invariance, variance, and covariance component. A weighted average is taken from these components to define the loss function. The first component is the variance component ($v(Z)$), where $Z$ is the batch:
\begin{equation}
    v(Z) = \frac{1}{d}\sum_{j=1}^{d}abs\big(\gamma - S(z^{j},\epsilon)\big),
\end{equation}
where $d$ is the number of features and $z^{j}$ is the specified feature value of all batch samples. $\gamma$ is a constant target value for the standard deviation, which is fixed at 1 as in the original paper. Finally $S(z^{j},\epsilon)$ is given by:
\begin{equation}
    S(x,\epsilon) = \sqrt{Var(x) + \epsilon},
\end{equation}
where $\epsilon$ is a small constant fixed to 0.0001. This function encourages the variance to be the same for all $j$ in the current batch. The covariance matrix of $Z$ is defined as:
\begin{equation}
    C(Z) = \frac{1}{n-1}\sum_{i=1}^{n}(z_{i}-\bar{z})(z_{i}-\bar{z})^{T},
\end{equation}
where
\begin{equation}
    \bar{z} = \frac{1}{n} \sum_{i=1}^{n}z_{i}.
\end{equation}
The regularization of the covariance can then be defined as:
\begin{equation}
    c(Z) = \frac{1}{d} \sum_{i\neq j}[C(Z)]_{i,j}^{2}.
\end{equation}
This term encourages the off-diagonal components to be close to 0. This decorrelates the variable of each embedding, which prevents informational collapse. The invariance component is given by the negative cosine similarity:
\begin{equation}
    s(Z_{i},Z_{i}') = -\frac{Z_i \cdot Z'_i}{\|Z_i\|\|Z'_i\|}.
\end{equation}

\begin{equation}
    l(Z,Z') = \lambda s(Z,Z') + \mu [v(Z) + v(Z')] + v[c(Z) + c(Z')].
\end{equation}
In this loss function, the scaling components are hyperparameters that need to be tuned.
% This function is very sensitive to the tuning of these hyperparameters. They need to be chosen carefully to prevent collapse. 
The overall objective function is given by \textcolor{black}{a summation over batches $I$ in dataset $D$ and over augmentations functions $t$ and $t'$ derived from the augmentation family $T$}:
\begin{equation}
    L = \sum_{I \in D} \sum_{t,t'~T} l(Z^{I}, Z'^{I}).
\end{equation}

\subsection{Parameter selection}
The proposed model is optimized for 30 epochs using a batch size of 2,048. Due to this large batch size, \textcolor{black}{the number of update steps for individual weights within a single epoch is reduced. This may be problematic if the number of epochs is not increased when the model is optimized by the commonly applied Adam optimizer. For this reason,} the model was optimized using the Layerwise Adaptive Rate Scaling (LARS) optimizer\citep{you2017large} with a learning rate of 0.01. \textcolor{black}{This optimizer updates layer-wise instead of the individual weights, reducing the need for more epochs during training.} A decay was implemented, where the learning rate is reduced by a factor of 10 if learning stagnates. \textcolor{black}{The weights of the loss function were set to $\lambda=5$, $\mu=5$, and $v=1$.} The complete pipeline was built using the PyTorch module and available on GitHub \footnote{\href{https://github.com/hildeingvildhummel/UATR}{https://github.com/hildeingvildhummel/UATR}}. For classification, a logistic regression was used, using the learned embeddings as features. \textcolor{black}{Finally, the performance of this classifier is evaluated using the accuracy score.}

\subsection{Baselines}
The proposed unsupervised method was compared to a supervised CL method inspired by the SimCLR framework \citep{uid} in terms of accuracy and a F1 score weighted by the support of the class. This baseline consists of a ResNet18 encoder followed by a two-layer MLP of dimension 128 as the projector. A smaller baseline architecture is chosen to cope with the limited quantity of labeled data. 

The baseline model is trained using the supervised contrastive loss function \citep{khosla2020supervised} \textcolor{black}{This loss function is derived from the original NTXent loss \citep{uid} which takes the following form:}

\begin{equation}
    \mathcal{L}^{self} = \sum_{i \in I}\mathcal{L}_{i}^{self} = - \sum_{i \in I}\log \frac{\exp (\mathbf{Z_i} \cdot \mathbf{Z_j} / \tau)}{\sum_{a \in \mathcal{I} \backslash \{i\}} \exp (\mathbf{Z_i} \cdot \mathbf{Z_a}/ \tau)},
\end{equation}
\textcolor{black}{in this function, let $i \in \mathcal{I} \equiv \{1..2N\}$ be the index of the sample of interest called the anchor sample and $j$ be it's augmented version and defined as the positive sample. All the other $2(N-1)$ samples are defined as the negatives and the temperature scalar parameter is defined as $\tau \in \mathcal{R}^+$.}

\textcolor{black}{This original function is rewritten to incorporate label information into the following format:}

\begin{equation}
    \mathcal{L}_{sup} = \sum_{i \in \mathcal{I}}\frac{-1}{|\mathcal{P}(i)|} \sum_{p \in \mathcal{P}(i)} \log \frac{\exp (\mathbf{Z_i \cdot \mathbf{Z_p}/\tau)}}{\sum_{a \in \mathcal{I} \backslash \{i\}} \exp(\mathbf{Z_i} \cdot \mathbf{Z_a} / \tau)}.
\end{equation}

\textcolor{black}{Here, $\mathcal{P}(i)$ is the set of indices for data samples with the same label as anchor $i$. The loss is summed over all anchor samples within the batch and normalized by $\mathcal{P}(i)$ which accounts for the number of positive pairs for each anchor.} This function aims to place the samples with the same label close together in the embedding space and pushes samples with other labels further away. The supervised baseline model \textcolor{black}{is trained on the labeled Deepship data and} is optimized using either the LARS optimizer with a learning rate of 0.01 or the Adam optimizer with a learning rate of 0.0005. \textcolor{black}{The Adam optimizer is also implemented since this optimizer is suggested in previous work that proposed a supervised CL framework for UATR \citep{nie2023contrastive, sun2023underwater, xie2023guiding}}

\section{Results}
\begin{figure}
    \centering
    \includegraphics[width=0.9\linewidth]{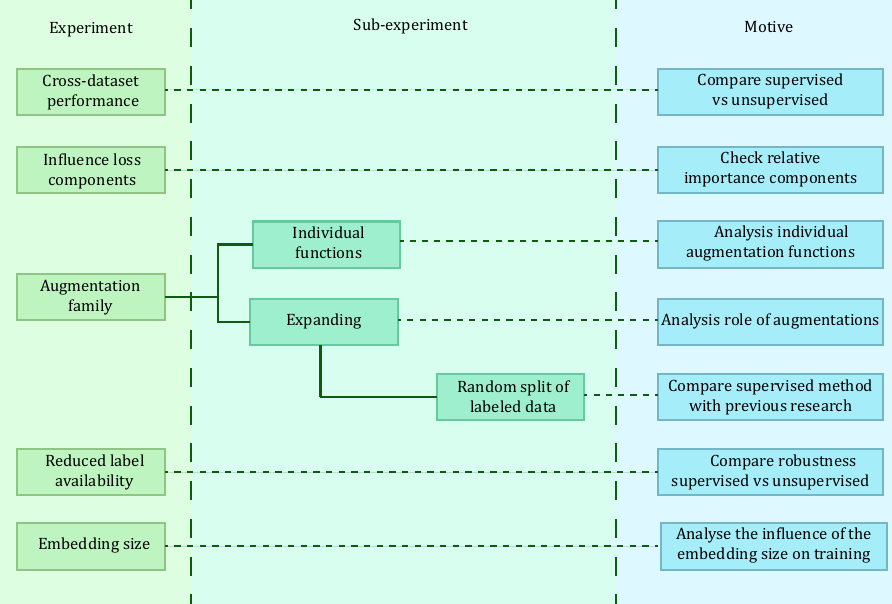}
    \caption{Overview of the experiments.}
    \label{fig:Experiments}
\end{figure}

\textcolor{black}{In this section, the generalizability of the proposed unsupervised method is evaluated and compared to the supervised CL framework. One way to estimate this is to measure cross-dataset performance. In this study, the datasets cover real-world recordings from various regions of the world with different ocean environments. In addition, an in-depth analysis of the weights of the VICReg loss components, augmentation functions, availability of labeled data, and embedding dimensions is presented. An overview of the experiments is visualized in Figure \ref{fig:Experiments}.}

\subsection{Multi-Purpose Classification}

\begin{table}[h]
\centering
\caption{The accuracy and weighted F1 score over multiple benchmark datasets.}
\label{tab:AccuracyGeneral}
\begin{tabular}{||l || c c | c c | c c||} 
 \hline
 Dataset & \multicolumn{2}{c|}{Deepship} & \multicolumn{2}{c|}{ShipsEar} & \multicolumn{2}{c||}{Watkin's} \\ [0.5ex]
 \hline
 & Accuracy & F1 & Accuracy & F1 & Accuracy & F1 \\
 \hline
 Supervised CL ResNet18 (Adam) & 53.94\% & 0.49 & 21.40\% & 0.08 & 54.54\% & 0.54\\
 Supervised CL ResNet18 (LARS) & \textbf{56.51\%} & \textbf{0.55} & 43.94\% & 0.42 & 80.46\% & 0.77 \\

  Supervised Conformer (LARS) & 55.16\% & 0.54 & 48.47\% & 0.49 & \textbf{87.10\%} & \textbf{0.85}\\

\hline
 Unsupervised ResNet18 (LARS) & 50.34\% & 0.49 & 42.25\% & 0.37 & 71.80\% & 0.63 \\
 Unsupervised Conformer (LARS) & 54.87\% & 0.54 & \textbf{57.42\%} & \textbf{0.57} & 86.10\% & 0.84\\
 \hline
\end{tabular}
\end{table}

First, performance was compared between supervised contrastive loss baseline models and unsupervised VICReg models. The supervised and unsupervised approaches were trained with a ResNet18 model and a Conformer model. The accuracy and weighted F1-score for the logistic regression models built on top of the backbones are given in Table~\ref{tab:AccuracyGeneral}. Note that the ResNet18 backbone model trained with supervised CL was optimized with both the Adam optimizer and the LARS optimizer. As mentioned previously, the Adam optimizer is still commonly used when building supervised CL frameworks for UATR. First, the results indicate that the use of a LARS optimizer improves the downstream performance of Deepship, which is also the dataset both backbones were trained on. Inclusion of the LARS optimizer also improves the generalizability of these backbone models to ShipsEar and Watkin's. In light of these results, corroborated by the fact that LARS-based optimizers are the standard when working with very large batch sizes, all further backbones were optimized using the LARS optimizer.

Further comparison is made between the ResNet18 model and the Conformer model. As the Conformer model has more parameters and is built specifically for audio, it is expected to perform better than the ResNet18 model. Although this hypothesis holds for the unsupervised approach, one surprising result is shown for the supervised CL ResNet18 on Deepship. Here, the ResNet18 supervised CL approach achieves the best performance in Deepship, even compared to the Conformer supervised CL approach. However, the supervised CL Conformer translates better to other datasets. While the ResNet18 supervised CL model is more capable of capturing the properties important for Deepship classification, the Conformer has learned a more general backbone model.

Finally, the difference between using a form of supervised CL and a form of unsupervised CL, is compared. For ResNet18, using supervised CL on Deepship builds both a better backbone for Deepship and a model more generalizable to other datasets. However, the trade-off becomes a bit more nuanced when comparing the Conformer models. The supervised CL Conformer model is clearly more capable on Deepship. This result is not surprising, as its backbone was trained on Deepship and thus already optimized to discriminate Deepship classes. On the other hand, the unsupervised Conformer achieves a similar performance on Watkin's and a better performance on ShipsEar. The features learned by the unsupervised CL approach thus seem more generalizable to other datasets. When the model capacity becomes larger, the use of a dataset such as Deepship to train a backbone model seems to bias it towards that dataset. 

\subsection{Influence Individual Loss Components of VICReg}
In the original VICReg paper \citep{bardes2021vicreg}, the authors demonstrated that at least the invariance and the variance components are essential to prevent collapse during training. They originally recommended a weight combination of $\lambda = 25$, $\mu = 25$, and $v = 1$, focusing on computer vision tasks. However, images encounter more contrast than spectrograms, making UATR more susceptible to \textit{informational collapse} \textcolor{black}{in which the variables of the embeddings carry redundant information \citep{bardes2021vicreg}}. Consequently, the covariance component of the VICReg loss function can be more important for underwater acoustic data than for computer vision-related tasks. For this reason, various weight combinations that were recommended in the Appendix of \citep{bardes2021vicreg} were tested (see Table \ref{tab:LossWeightInfluence}). This confirmed that the relative importance of the covariance component needs to be increased to optimize the embeddings in UATR.

\begin{table}[!ht]
\caption{The influence of the weights assigned to the individual loss components of VICReg on the classification of Ship types.}
\label{tab:LossWeightInfluence}
\begin{center}
\begin{tabular}{||c c c || c||} 
 \hline
 $\lambda$ & $\mu$ & $v$ & Accuracy \\ [0.5ex]
 \hline\hline
1 & 1 & 1 &  55.35\%\\
5 & 5 & 1 & \textbf{55.61\%} \\
25 & 25 & 1 & 23.81\% \\
 \hline
\end{tabular}
\end{center}
\end{table}

\subsection{Influence of the Augmentation Family}
\textcolor{black}{An augmentation family, consisting of various augmentation functions, is needed to define positive samples for unsupervised CL training. During training, a single audio recording is augmented by randomly selecting two individual functions from this family. Then both augmented samples will be optimized to be close together in the expanded space using CL. }The performance of a model optimized using CL, may be sensitive to the choice of the augmentation functions. For this reason, the influence of the different augmentation functions within the specified family is analyzed (see Table \ref{tab:AugmentationInfluence}). The highest accuracy scores on both Deepship and Watkin's are achieved by the LowPass filter. This augmentation function alone achieves performance similar to that of the model trained utilizing the full augmentation family.  This is likely due to the fact that the most discriminative information in the spectrogram is primarily located in the lower frequency bands. This result indicates that the unsupervised model focuses on the most informative spectral regions. However, the performance of the model solely using the LowPass filter is reduced in ShipsEar. This suggests that the combination of all the augmentation functions is needed to achieve stable cross-dataset performance. 

\begin{table}[!ht]
\caption{The influence of the augmentation functions on the classification of Ship types.}
\label{tab:AugmentationInfluence}
\begin{center}
\begin{tabular}{||l || c c | c c | c c ||} 
 \hline
 Augmentation Function & \multicolumn{2}{c|}{Deepship} & \multicolumn{2}{c|}{ShipsEar} & \multicolumn{2}{c||}{Watkin's} \\ [0.5ex]
 \hline\hline
 & Accuracy & F1 & Accuracy & F1 & Accuracy & F1 \\
 \hline
LowPass & \textbf{55.93\%} & 0.54 & 49.42\% & 0.44 & \textbf{87.00\%} & 0.83\\
Noise & 
52.54\% & 0.52 & 
50.19\% & 0.48 & 
85.21\%& 0.83\\
MixUp & 
55.25\% & 0.54 & 
51.86\% & \textbf{0.51} & 
86.03\% & \textbf{0.84}\\
\hline
Combined & 55.61\% & \textbf{0.55} & \textbf{52.14\%} & 0.50 &  \textbf{87.00\%} & 0.83\\
 \hline
\end{tabular}
\end{center}
\end{table}

\subsubsection{Expanding the Augmentation Family}
The augmentation family was expanded by multiple speech-based augmentation functions, such as varying gain, reverb, pitch, and performing polarity inversion. However, introducing more types of speech-wise augmentations does not improve the unsupervised model performance on all three datasets (see Table \ref{tab:FullAugmentation}). The performance of the unsupervised conformer on Deepship is improved; however, the performance on the other datasets is decreased. This shows the importance of representative augmentation functions in guiding the model in an unsupervised learning framework. In contrast, in the supervised setting, these models benefited from the additional augmentation functions. Here, both models show an increase in performance on all three datasets. The performance is greatly improved, \textcolor{black}{showing that the generalization capacity of the supervised models is increased by expanding the augmentation family. The supervised ResNet18 model trained with the LARS optimizer has a competitive performance similar to the unsupervised learning model with the original augmentation family. Expanding the augmentation family increases the variability of the data during training, since the supervised method can rely on label information, this method can benefit from this additional variability. This is not the case for the unsupervised method, which does not depend on high-quality labeled data.}

\begin{table}[!ht]
\caption{The performance of both the Supervised and Unsupervised frameworks with an expanded augmentation family, incorporating speech based augmentation functions.}
\label{tab:FullAugmentation}
\begin{center}
\begin{tabular}{||l || c c | c c | c c||} 
 \hline
 Model & \multicolumn{2}{c|}{Deepship} & \multicolumn{2}{c|}{ShipsEar} & \multicolumn{2}{c||}{Watkin's} \\ [0.5ex]
 \hline\hline
 & Accuracy & F1 & Accuracy & F1 & Accuracy & F1 \\
 \hline
 Supervised CL ResNet18 (Adam) & \textbf{63.07\%} & \textbf{0.63} & 45.91\% & 0.43 & 75.73\% & 0.71\\
Supervised CL ResNet18 (LARS) & 61.17\% & 0.60 & \textbf{52.64\%} & \textbf{0.49} & 86.33\% & 0.84\\
Supervised CL Conformer (LARS) & 56.55\% & 0.56 & 45.41\% & 0.44 & \textbf{86.94\%} & \textbf{0.85}\\
\hline
Unsupervised CL ResNet 18 (LARS) & 42.48\% & 0.39 & 38.24\% & 0.33 & 69.66\% & 0.62 \\
Unsupervised CL Conformer (LARS) & 57.78\% & 0.56 & 50.47\% & 0.48 & 86.14\% & 0.84\\
 \hline
\end{tabular}
\end{center}
\end{table}

\subsubsection{Random Data Split for the Supervised Model}
\begin{figure}
    \centering
    \includegraphics[width=1.0\linewidth]{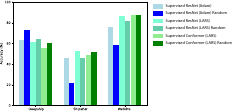}
    \caption{Accuracy of supervised baseline models trained on either time-wise split or random split of Deepship.}
    \label{fig:FairvsRandom}
\end{figure}

\begin{figure}
    \centering
    \includegraphics[width=0.9\linewidth]{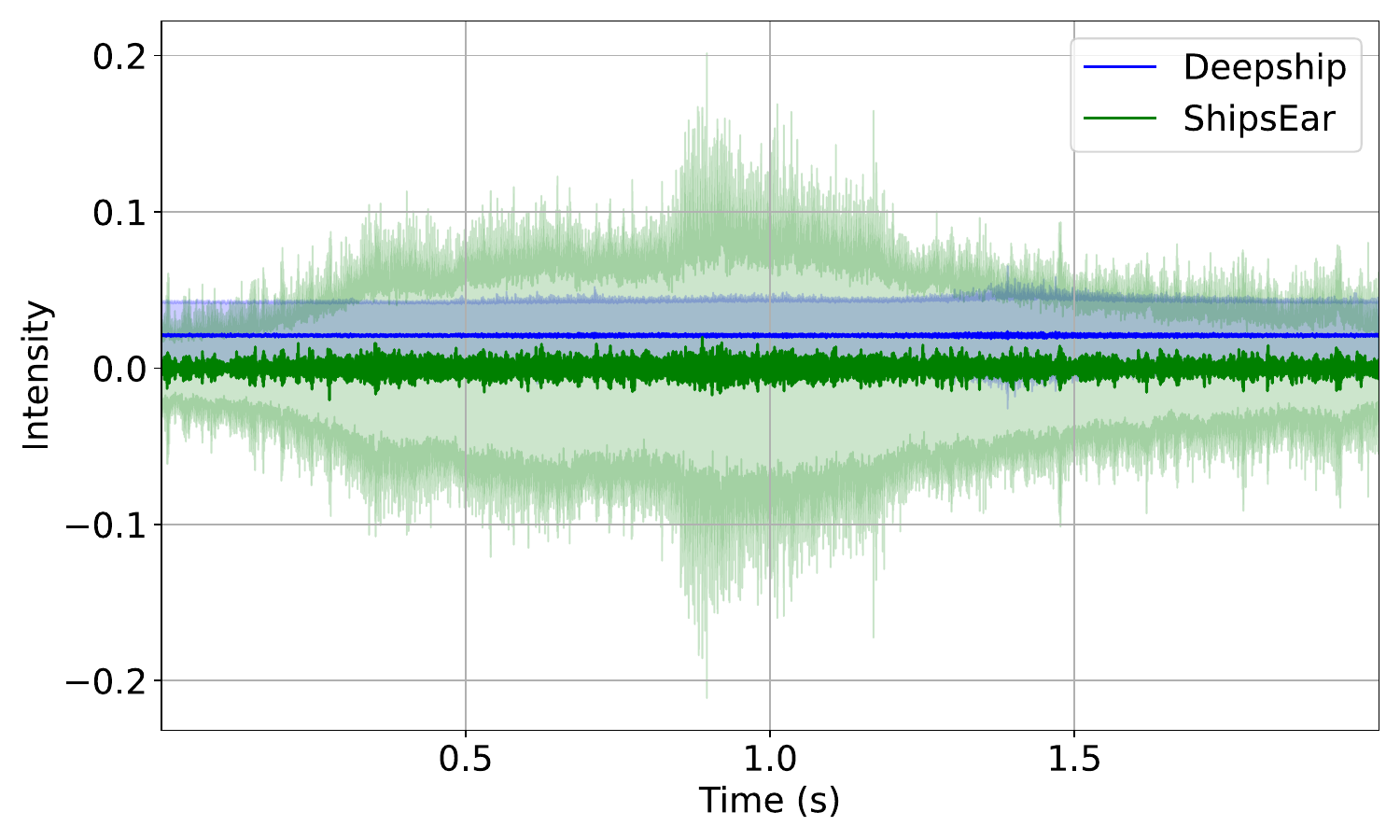}
    \caption{Temporal Variation of Deepship and ShipsEar, showing the mean and the \textcolor{black}{variability}.}
    \label{fig:temporalVariance}
\end{figure}

So far, the models are evaluated using a time-wise split for training and test set. However, previous research on automatic ship recognition adopted a random split for evaluation \citep{hummel2024survey}. \textcolor{black}{Unfortunately, such a split provides a poor reflection of the performance of the model in real-world conditions. }To examine the impact of the train-test split, a random split was generated by randomly splitting the individual Deepship recordings into a training set (80\%) and a test set (20\%), ensuring that no individual recordings were present in both sets. The baseline models were trained in a supervised manner using the expanded augmentation family. Next, the performance of the models is examined by translating them again to ShipsEar and Best of Watkin's (see Figure \ref{fig:FairvsRandom}). Here, both ResNet18 models have an increased accuracy score when trained on the Random split\textcolor{black}{, with a comparable performance to previous research \citep{hummel2024survey}}. This is because the number of newly seen ships is smaller compared to the time-wise split. In addition, the impact of the different seasons is reduced with the random split. However, the generalization performance for both ShipsEar and Best of Watkin's is worse when the ResNet18 is optimized on the random split. \textcolor{black}{An explanation for this could be that the Deepship data is more stationary compared to ShipsEar and Watkins, which capture more temporal fluctuations (see Figure \ref{fig:temporalVariance}).} On the other hand, the supervised conformer model achieves the lowest performance on Deepship for both the time-wise and the random split. This is probably due to the model size, the conformer model is significantly bigger compared to ResNet18 and therefore the Deepship dataset is probably too small to properly train the model. The performance of the model trained on the random spit is slightly higher on ShipsEar and has a similar on Watkin's. This shows that the defined split in the literature results in a higher reported accuracy on Deepship, however it may  \textcolor{black}{reduce} its generalization capacity in other related tasks or regions. 

\subsection{Impact of Reduced Labeled Data}
\begin{figure}
    \centering
    \includegraphics[width=0.9\linewidth]{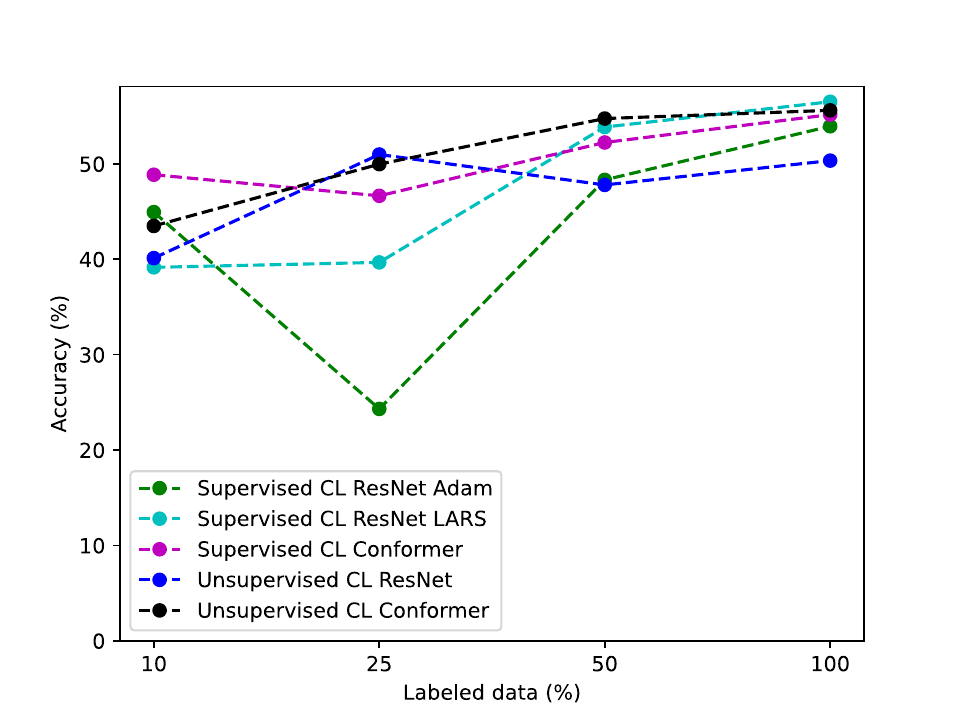}
    \caption{The accuracy on Deepship when trained on reduced amount of labeled data. }
    \label{fig:ReducedLabels}
\end{figure}

The performance of the supervised models is known to rely on a large amount of labeled data. To evaluate the impact of reducing labeled data on model performance, the size of the Deepship training dataset was reduced by randomly selecting a subset of individual recordings for training. The unlabeled data used for unsupervised CL remained the same. This experiment explores the relative impact of a reduced set of available labeled data for both methods. The impact on performance for both the baseline models and the unsupervised models are visualized in Figure~\ref{fig:ReducedLabels}. Here, the unsupervised Conformer method outperforms the supervised methods when the number of labeled data samples is reduced. This observation highlights the robustness of the unsupervised method compared to supervised models. A notable exception occurs at the 10\% dataset size, where the ResNet18 model trained with the Adam optimizer and the supervised Conformer exhibit better performance than the unsupervised model. This result is likely due to the redundancy of individual ships in the 10\% dataset compared to the 25\% dataset, creating a performance peak for this configuration.

\subsection{Influence of Embedding Size}
\begin{figure}
    \centering
    \includegraphics[width=0.9\linewidth]{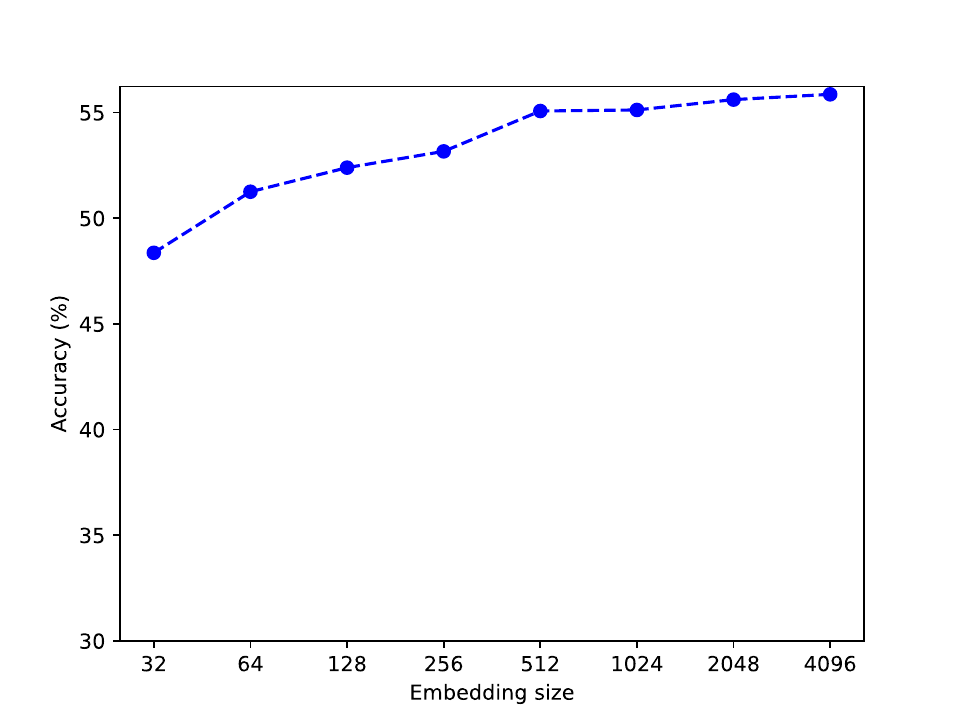}
    \caption{The accuracy of the unsupervised conformer model on Deepship over the embedding size.}
    \label{fig:embedding_size}
\end{figure}

The size of the embedding vector directly influences the ability of the model to learn the patterns within the dataset. Larger vectors have the ability to store more features; however, they increase the risk of underfitting the data and need more computational resources. However, smaller vectors may not capture essential patterns within the training dataset. This trade-off is examined in the proposed unsupervised framework, with the results illustrated in Figure \ref{fig:embedding_size}. Here, the \textcolor{black}{accuracy score increases} as the embedding size increases to 512, with only slight gains beyond this point. Further examination of the loss components during training showed that larger embedding sizes are associated with higher covariance loss and lower variance loss, while smaller sizes exhibit the opposite trend. These results align with the idea that larger vectors ease higher variations among them but also increase the likelihood of correlated elements, \textcolor{black}{making the model prone to \textit{informational collapse}}.

\subsection{Computational Complexity}

\begin{table}[!ht]
\caption{Number of parameters and inference time.}
\label{tab:CompComplexity}
\begin{center}
\begin{tabular}{||l || c c ||} 
 \hline
 & Parameter Count & Minimum Inference Time (ms) \\
 \hline
ResNet18 & 16,690,544 & 2.817 \\
Conformer & 101,177,392 & 5.268\\
\hline
\end{tabular}
\end{center}
\end{table}

To evaluate the cost of running these models, the parameter count and the mean inference time were recorded for both. The mean inference time was based on the minimum time to process a sample calculated over $10$ passes of the Deepship evaluation dataset. These numbers were recorded on a compute node consisting of 9 cores of an Intel Xeon GPU, 60GB of RAM memory, and an A100 40GB GPU. The training of all models took at most 10 hours under this setup.

Note that the minimum inference time was taken per sample. This is the standard choice when recording the runtime of a task. The minimum time is closest to the true inference time, as variance is usually caused by external factors, such as other running processes. Here, we observe that processing a sample for both models only takes a few milliseconds. Although the Conformer model is around six times larger than the ResNet18 model, its inference time is less than double. This shows that scaling to larger models is viable due to the high degree of parallelizability. While the setup is quite large, the low inference times indicate that a real-time edge solution may be viable as well. Both models can process a sample of two seconds within a couple of milliseconds, demonstrating the computational viability of AI-based approaches for UATR.

% First, it is important to note that the conformer model is \~6 times larger than the ResNet model. Despite this fact, the mean inference time is only \~1.5 times that of the ResNet models for both supervised CL and unsupervised CL. This is likely due to the parallelizability of the larger Conformer model. Another observation is that the mean inference time of the supervised approaches is less than half that of the unsupervised approaches. This can be largely explained by the absence of an additional set of positive samples. A second forward pass is performed for the positive samples for the unsupervised approach. For supervised CL, the positive pairs are formed within the batch based on additional label information provided. Note that this difference is thus largely caused by the specific implementation; the final backbone will have the same inference time. Overall, an interesting observation to make is the fact that even a Conformer model is potentially viable to use. A Conformer model can process 2048 samples of 2 seconds within 160ms. This demonstrates that large models can be viable within UATR.

\section{Discussion}
In this paper, a new proof-of-concept small architecture is proposed on unlabeled data. The model successfully extracted informative features from this unlabeled dataset and was able to transfer this knowledge to other labeled datasets by classifying both the sounds of marine mammals and the types of ships. In particular, the unsupervised model correctly classified newly seen ships into their corresponding class. The proposed unsupervised framework shows performances comparable to the supervised baselines. However, the performance of these supervised models is much more reliant on the amount of available labeled data. As the unsupervised framework does not require labeled data to extract informative features, its performance does not degrade as drastically when less labeled data is available.

The choice of augmentation functions proved important for the performance and generalizability of the backbone models. Using multiple augmentations tailored specifically to underwater acoustics resulted in a strong performance on all datasets for both supervised CL and unsupervised CL. The additional inclusion of speech based augmentation functions gave mixed results with a marked improvement in accuracy for the smaller supervised CL models. Furthermore, this work used a time-wise split as the standard split for Deepship, as opposed to the commonly used random split. Using a time-wise split reduced the accuracy and F1-scores on Deepship. However, this work argues that a time-wise split provides a more realistic view of the real-world performance of these models.

\textcolor{black}{The defined ship types in the labeled datasets have similar acoustic profiles with a high spread of acoustic signatures within a single class and a great overlap between classes. This makes the categorization of the acoustic profiles into these classes challenging and implies that a more suitable categorization of these profiles is needed. A solution to this problem could be to create classes based on the characteristics of the ship, such as the number of blades, the size of the ship, and the type of engine(s), instead of the first-level \textit{Automatic Identification System} (AIS) categorization. }

Overall, the proposed method achieves a performance similar to that presented in \citep{tian2023few}. However, they reported a larger framework for the model, and they implemented few-shot learning using only Deepship data treating part of the data as unlabeled and not by making the translation between two different datasets. An advantage of this unsupervised approach over few-shot learning is that there is a great amount of unlabeled data available. Therefore, this framework can be extended by incorporating more diverse underwater acoustical data from multiple publicly available hydrophones. The results demonstrate that large datasets require models with more capacity to achieve sufficient generalization. In addition, the runtime analysis demonstrates that larger models do not necessarily pose issues when it comes to inference time, as processing a 2-second sample takes milliseconds. By incorporating more data, this framework could translate better to other underwater acoustic analysis tasks, such as climate change monitoring and nuclear bomb test detection.

\section{Conclusion}
This work highlights the potential of unsupervised CL methods in UATR. It explored the implementation of abundantly available unlabeled lower-quality underwater acoustic data and presented a CL pipeline for robust generalized embedding generation with competitive performance to supervised models. Overall, this shows the potential for the possibilities of unsupervised learning methods in the automatic analysis of large amounts of passive acoustic monitoring recordings. The proposed unsupervised CL has shown to be generalizable across multiple datasets and related underwater acoustic tasks. This makes the pipeline extendable for multiple automatic underwater acoustic analysis tasks. This creates the possibility to automatically identify and quantify sound pollution without depending on a large volume of high-quality labeled data.

\hfill

{\bf Declaration of Competing Interest:} None of the authors has any relationship with people or organizations that can influence their work.

\bibliographystyle{elsarticle-harv}\biboptions{authoryear}
\bibliography{references}

\end{document}